\input epsf
\newcommand{\be}{\begin{equation}}
\newcommand{\ee}{\end{equation}}
\newcommand{\bea}{\begin{eqnarray}}
\newcommand{\eea}{\end{eqnarray}}
\newcommand{\bean}{\begin{eqnarray*}}

\newcommand{\eean}{\end{eqnarray*}}
\font\upright=cmu10 scaled\magstep1
\font\sans=cmss10
\newcommand{\ssf}{\sans}
\newcommand{\stroke}{\vrule height8pt width0.4pt depth-0.1pt}
\newcommand{\Z}{\hbox{\upright\rlap{\ssf Z}\kern 2.7pt {\ssf Z}}}

\newcommand{\C}{{\rlap{\rlap{C}\kern 3.8pt\stroke}\phantom{C}}}
\newcommand{\R}{\hbox{\upright\rlap{I}\kern 1.7pt R}}
\newcommand{\CP}{\C{\upright\rlap{I}\kern 1.5pt P}}
\newcommand{\PP}{\hbox{\upright\rlap{I}\kern 1.5pt P}}

\newcommand{\identity}{{\upright\rlap{1}\kern 2.0pt 1}}

\newcommand{\taubf}{\mbox{\boldmath $\tau$}}
\newcommand{\bn}{{\bf n}}

\def\sketch_hopf.eps{1}
\def\initial.ps{2}
\def\one.ps{3}
\def\two.ps{4}
\def\three.ps{5}
\def\pos.ps{6}
\def\hopf.ps{7}
\def\ene.ps{8}
\def\tori.ps{9}
\def\energy_graph.ps{10}
\def\recon.ps{11}
\def\sixmov.ps{12}
\def\sevenmov.ps{13}
\def\eightmov.ps{14}

\documentstyle[12pt,a4wide]{article}

\newcommand{\news}{\setcounter{equation}{0}}
\begin{document}
\title{\vskip -70pt
\begin{flushright}
{\normalsize UKC/IMS/98-34} \\
{\normalsize DAMTP-1998-110} \\
\end{flushright}\vskip 50pt
{\bf \LARGE \bf Solitons, Links and Knots}} 
\author{Richard A. Battye$^{\ \dagger}$ and Paul M. Sutcliffe$^{\ \ddagger}$\\[10pt]
\\{\normalsize $\dagger${\sl Department of Applied Mathematics and Theoretical Physics,}}
\\{\normalsize{\sl University of Cambridge,}}
\\{\normalsize {\sl Silver Street, Cambridge, CB3 9EW, U.K.}}
\\{\normalsize {\sl Email : R.A.Battye@damtp.cam.ac.uk}}\\ 
\\{\normalsize $\ddagger$ {\sl Institute of Mathematics, University of Kent at Canterbury,}}\\
{\normalsize {\sl Canterbury, CT2 7NZ, U.K.}}\\
{\normalsize{\sl Email : P.M.Sutcliffe@ukc.ac.uk}}\\}
%\date{}
\maketitle
\vskip 25pt
\begin{abstract}
Using numerical simulations of the full nonlinear equations of motion
we investigate topological
solitons of a modified $O(3)$ sigma model in three space dimensions, in
 which the solitons are stabilized by the Hopf charge.
We find that for solitons up to charge five the solutions have
the structure of closed strings, which become increasingly twisted as
the charge increases. However, for higher charge the solutions are
more exotic and comprise linked loops and knots.
We discuss the structure and formation of these solitons and demonstrate
that the key property responsible for producing such a rich variety of solitons
is that of string reconnection.\\

\end{abstract} 
\newpage

\section{Introduction}\news

Topological solitons are of great interest in a number of areas, including particle physics,
 cosmology and condensed matter physics. However,  only in the last few years, with the advances
 in computing power, has it become possible to fully investigate topological solitons in three
 spatial dimensions. Recent results on Skyrmions~\cite{BS1,BS2,BS3} and monopoles (see for example,
 ref.~\cite{monopole}) have revealed 
that intricate and fascinating structures appear, which are inherently three-dimensional and 
therefore cannot arise as generalizations of solitons in one or two space dimensions. Therefore,
 it is of considerable interest to investigate other three-dimensional soliton models, in order 
to determine the type of behaviour we may expect to see in this rapidly developing field.

In the majority of topological soliton models that have been studied
 to date the topological charge, or soliton number, arises as a
 winding number between spheres of equal dimension.  For example, in
 gauged $SU(2)$ models the monopole number is the degree of a map
 between 2-spheres, which counts the winding of the Higgs field in the
 gauge orbit of vacua at spatial infinity. Similarly, in the Skyrme
 model the baryon number is a winding number between two 3-spheres,
 which counts the wrapping of the $SU(2)$ Skyrme field as a map from
 compactified space. By contrast, in the modified  $O(3)$ sigma model
 we consider here the topological characteristic responsible for the
 stability of the solitons is not a winding number, but is a linking
 number of the field lines. There is an associated map between
 spheres, but the spheres do not have the same dimension ($S^3$ to
 $S^2$) and the topological charge is the Hopf  invariant of this
 map. From the point of view of identifying the soliton, these
 solitons will be line-like, as opposed to point-like as is the case
 for monopoles and Skyrmions, indicating that some very different
 solitonic structures are likely. 

There have been two preliminary numerical investigations~\cite{FN,GH} of solitons in this model
 with charges one and two. Both concentrated mainly on axially symmetric configurations, using 
toroidal or cylindrical coordinates so that the numerical problem is effectively reduced to one in two 
spatial dimensions. Although this approach has the advantage of substantially reducing the resources 
required to relax configurations, it does suffer from two serious drawbacks which make 
interpretation of the results difficult. Firstly, only axially symmetric solitons can be 
studied, which is unlikely to be appropriate for   minimum energy multi-solitons of arbitrary
 charge. The second, more technical problem is that in these coordinate systems, the
 complicated topological nature of the solutions requires the imposition of subtle boundary
 and regularity conditions. Not surprizingly, some of the results of these investigations,
 such as the distribution of the energy density for the charge one solution, are qualitatively
 different due to differing choices of these conditions.

In the work reported here we avoid these difficulties by performing fully three-dimen\-sional
 simulations in Cartesian coordinates, where such problems simply do not arise; of course, 
the price to be paid for this is a substantial increase in the required computer power. But
 given sufficient resources (for example, a parallel super-computer), this approach is the 
most conservative discretization of the model. The code used is a modified version of that 
developed to investigate the dynamics and bound states of Skyrmions in (3+1) dimensions 
(see ref.\cite{BS3} for a detailed description). It has been substantially tested and 
produced some very attractive results in that context. When applied to the 
modified $O(3)$ sigma model,
 we reproduce the common results of refs.~\cite{FN,GH}, but where differences arise, we find
 agreement with the work of Gladikowski and Hellmund~\cite{GH}. This suggests two things: (i)
 the assumption of axial symmetry is sufficient for charges one and two, (ii) the imposition 
of the boundary and regularity conditions was done incorrectly in ref.\cite{FN}.

The results presented in the subsequent sections of this paper (see
 also ref.~\cite{BS5}) go much further than resolution
 of the differences between refs.~\cite{FN,GH}. We present candidate minimum energy configurations
 for charges one to eight which suggest that exotic linked and knotted solitons may also exist at
 higher charges. This is in keeping with the spirit of ref.~\cite{FN}, where it was suggested that
 a knotted soliton is possible even at charge three. However, the details are very much different.
 The charge three soliton is just a twisted torus, as are those for charge four and five, with the
 degree of twisting becoming more noticeable as the charge increases. But at charge six the 
solution is very different; it being constructed from two linked tori. At charge seven the 
solution resembles that of a trefoil knot and at charge eight it comprizes of two tori which 
are linked twice. We will discuss the precise details of each soliton and attempt to identify 
a principle dictating the structure of them, as we did for Skyrmions~\cite{BS2,BS3}. This is more
 difficult in this case since as we have already noted the solitons are line-like, as opposed
 to point-like, and hence the solutions do not have point symmetries. We believe, nonetheless,
 that we have identified some interesting qualitative trends.

The identification of these complex solitonic structures is interesting in its own right as a
 mathematical exercise. However, we believe that the model in question also has some important
 physical implications in condensed matter and particle physics. Many condensed matter
 systems are described by an $O(3)$ unit vector order parameter, examples being nematic liquid 
crystals and magnetic bubbles in electron gas systems, and hence we have identified candidate
 solitonic structures which may be found experimentally. More speculatively,  Faddeev and Niemi
 have suggested that this model could be used as an approximation for strongly coupled $SU(2)$
 Yang-Mills theory~\cite{FN2}, in which case the solitons would be confined glueballs. Although
 we will not refer further to these or any other possible applications, it is important to 
realize that soliton models have been found to have many applications in a variety of physical
 contexts. 

\section{The modified $O(3)$ sigma model}\news

Some time ago Faddeev~\cite{Fa} suggested that stable closed strings may exist as topological 
solitons in an $O(3)$ sigma model which included a fourth order derivative term, with the topology
 arising due to the twisting of a planar soliton as it is embedded in three-dimensional space.
 Explicitly, 
the field of the model is a real three-component vector ${\bn}=(n_1,n_2,n_3)$, with unit length
 ${\bf n}\cdot{\bf n}=1,$ and the Lagrangian density in 
(3+1)-dimensions is given by
\be
{\cal L}=\partial_\mu \bn\cdot \partial^\mu \bn
-\frac{1}{2}(\partial_\mu \bn \times \partial_\nu \bn)\cdot
(\partial^\mu \bn \times \partial^\nu \bn)\,.
\label{lag}
\ee
The first term in the Lagrangian (\ref{lag}) is that
of the $O(3)$ sigma model and the higher order derivative 
term is required to prevent an instability of configurations
under a rescaling of the space coordinates, that is, it is a Skyrme-type term.
In fact, the usual Skyrme model~\cite{Sk}, in which the field takes values on a three-sphere, 
may be consistently restricted to a two-sphere equator and this exactly reproduces the model
 (\ref{lag}).

In order for a solution to have finite energy the vector $\bn$ must tend to a constant value at 
spatial infinity, which we take to be given by
$\bn_\infty=(0,0,1).$ The novel aspect of the model under consideration here is that if we restrict
 to finite energy configurations, then they have a topological characterization. The boundary 
condition implies that
space is compactified from $\R^3$ to $S^3$ so that at any fixed time the field is a map 
$\bn: S^3 \mapsto S^2.$ Since $\pi_3(S^2)=\Z$,  each field has an associated integer
 topological charge $Q$, the Hopf charge, which gives the soliton number.

As already discussed the soliton number is not a simple winding number, like it is for other
 solitons such as Skyrmions or monopoles, but rather it is a linking number between field lines.
 Formally, let $\omega$ denote the area two-form on the target $S^2$ and let $F=\bn^*\omega$ be 
its pullback under $\bn$ to the
domain $S^3.$ Then due to the triviality of the second cohomology group
of the 3-sphere, $H^2(S^3)=0$, this pullback must be exact, say
$F=dA,$ and hence the Hopf charge can be constructed by integrating  the Chern-Simons term over $\R^3$,
\be
Q=\frac{1}{4\pi^2}\int d^3{x}\,F\wedge A\,.
\ee
An important point to note is that in general the Hopf charge can not
be written as the integral of a density which is local in the field $\bn,$ a fact which will have 
some practical implications for our numerical results.
For this reason it is useful to consider a heuristic 
interpretation of $Q.$ The preimage of a point on the target $S^2$
is a closed loop. Now if a field has Hopf number $Q$ then the two loops consisting of the
 preimages of any two distinct points on the target $S^2$ will be linked exactly $Q$ times.
 Later we shall use this description, in terms of the linking number of field lines, to identify
 the Hopf number of various field configurations. In fig.~\sketch_hopf.eps 
we schematically represent the
 preimages of two distinct points in the target space for a configuration with $Q=1$.

Recall that the position ${\bf x}_0$ of a soliton is usually defined to be
the point in space at which the field takes the value
antipodal to the vacuum value, which in this case gives $\bn ({\bf x}_0)=-\bn_\infty.$ Thus 
the position of the soliton is the curve in space
described by the preimage of the vector $(0,0,-1).$ 
We shall see that displaying this closed string, giving the position of the
soliton, is a useful way to represent the solution.

Not only is there a topological charge in this model, but also a lower bound
on the energy in terms of the charge exists~\cite{KR}. Explicitly
the bound on the energy $E$ of a configuration with Hopf charge $Q$ is
\be E\ge c\vert Q\vert ^\frac{3}{4} \label{bound} \ee
where $c$ is the constant $c=16\pi^23^\frac{3}{8}\approx 238.$
Note that this energy bound is rather unusual in that a fractional
power of the topological charge occurs, reflecting the fact that
this bound is not obtained from the usual Bogomolny-type argument, but
relies on a sophisticated piece of analysis for its derivation.
As we shall see later, in comparing the above bound to the energy of
the computed solitons we conclude that the fractional power dependence
fits quite well, although the overall normalization constant $c$ could probably
be dramatically increased to obtain a much tighter bound.

The existence of a conserved charge and a lower bound on the energy, plus the stability to scaling,
 suggests --- but does not prove --- the existence of finite energy configurations at each charge.
 In order to proceed further the equations of motion for the field $\bn$ must be solved, but this
 is analytically intractable even in the case of $Q=1$. Hence, a numerical approach is 
required. The particular numerical scheme used  is a modification of our algorithm constructed to study
 Skyrmions~\cite{BS1,BS2,BS3}, whose salient features are a 
finite difference scheme on a grid containing $100^3$ points, where spatial derivatives are 
approximated by fourth-order accurate finite differences, and time evolution via a second-order
 leapfrog method.  In order to relax to static solutions from suitably 
random initial conditions, we use a number of techniques including the addition of a
 dissipative term to the equations of motion and periodic removable of all kinetic energy 
in the system. All the details of the numerical algorithm and 
the relaxation procedure are described in ref.~\cite{BS3}.

The methods we employ are the most simple and general discretization of the equations of motion.
 They make no assumption as to the spatial symmetry of the solution or its structure at the origin 
and hence provide an obvious improvement on the methods employed
 in refs.~\cite{FN,GH}. The only extra ingredient required is the imposition of a
 boundary condition, which can be done simply at the edge of the box, mimicking spatial infinity.

\section{Initial Conditions}\news

Once a numerical code has been implemented the next stage is to provide suitable initial
 conditions for a given charge, which can then be relaxed to yield the static soliton solutions. 
Thus, we need
 to be able to write down various 
configurations in which the field lines have the correct topological linking  
structure for the Hopf charge, under consideration. In refs.~\cite{FN,GH} configurations of
 charge one and two were produced by employing a toroidal or cylindrical ansatz. Here, we
 describe how a wide range of fields can be obtained for any Hopf charge without the need 
to refer to a particular coordinate system 
which is tailored to a given configuration.

We begin with the observation~\cite{Me} that a field with Hopf
charge $Q$ can be obtained by applying the standard Hopf map to
a map between 3-spheres which has winding number $Q.$ More explicitly, let $U({\bf x})$ be
 a Skyrme field, that is, any smooth map from $\R^3$ into $SU(2)$ which satisfies the boundary 
condition that $U$ tends to the
identity matrix as $\vert {\bf x}\vert\rightarrow\infty.$ Since the group
manifold of $SU(2)$ is $S^3$ and the boundary condition compactifies space,
then effectively we have that $U$ is a map between 3-spheres. Let $B$
denote the winding number of this map, which is known as the topological charge or baryon number
of the Skyrme field. Writing the components of $U$ in terms of complex numbers $Z_0$ and $Z_1$ as
\be
U=\left(\begin{array}{cc}Z_0&-\bar Z_1\\
Z_1&\bar Z_0\end{array}\right)\,,
\ \ \ \ 
\mbox{where}
\ \ 
\vert Z_0\vert^2+\vert Z_1\vert^2=1\,,
\ee
then the standard Hopf map can be written in terms of the column vector
$Z=(Z_0,Z_1)^T$ as
\be
\bn =Z^\dagger\taubf Z\,,
\label{hmap}
\ee
where $\tau_j$ denote the Pauli matrices. It is easy to see that the vector
defined by (\ref{hmap}) has unit length and satisfies the boundary condition
$\bn(\infty)=\bn_\infty.$ Furthermore, it can be shown that $Q=B$, that is, the
Hopf charge of the configuration constructed in this way is given by
the topological charge of the Skyrme field used.
Therefore, it is possible to construct field configurations
with non-trivial Hopf charge given a suitable supply of Skyrme fields.
Fortunately some recent work~\cite{HMS} has provided a method for constructing
Skyrme fields from rational maps between Riemann spheres and so we can apply
these techniques.

The first step is to write the Skyrme field in terms of a profile function
$f$ and a direction in the $su(2)$ algebra determined by a unit 3-vector
${\bf v}=(v_1,v_2,v_3).$ The explicit relation is simply
\be U=\exp(if{\bf v}\cdot\taubf). \ee 
When combined with the Hopf map (\ref{hmap}) this results in the fields
\bea
n_1&=&2(v_3v_1\sin f-v_2\cos f)\sin f\label{n1}\,,\\
n_2&=&2(v_3v_2\sin f+v_1\cos f)\sin f\label{n2}\,,\\
n_3&=&1-2(1-v_3^2)\sin^2 f\,,
\label{n3}
\eea
which parameterizes the unit vector $\bn$ in terms of another unit vector ${\bf v}$ and the profile
 function $f$.

Now, we follow the approach used for Skyrmions~\cite{HMS} and decompose the Skyrme
field into a radial and an angular dependence. Introducing the usual spherical
polar coordinates $(r,\theta,\phi)$, the ansatz assumes that the profile function depends only on the 
radial coordinate, that is, $f(r)$ and that the direction in the algebra determined by ${\bf v}$ is 
independent of $r$. The boundary conditions are $f(0)=\pi$ and $f(\infty)=0$, with $f(r)$ a 
monotonically decreasing function. The details of this function are not important, since we are only 
interested in creating initial conditions for relaxation, so that any reasonable smooth function
 with the correct boundary conditions 
 will suffice.

Note that the unit vector ${\bf v(\theta,\phi)}$ is now a map between 2-spheres,
and therefore has a winding number. This winding number is precisely the baryon
number $B$ of the Skyrme field and hence also the Hopf charge $Q$ of the configuration we generate. 
To prescribe this map between 2-spheres it is convenient to use a Riemann sphere parameterization
 of both the 2-spheres. Explicitly, we use the  coordinate $z=e^{i\phi}\tan(\theta/2)$ obtained by
 stereographic projection and similarly for the target sphere, where we represent a point as a
 complex number $R(z)$ via
\be
{\bf v}=\frac{1}{1+\vert R\vert^2}(R+\bar R,i(\bar R-R),\vert R\vert^2-1)\,.
\label{vr}
\ee
We can now generate configurations of Hopf charge $Q$ by taking $R(z)$ to be a
rational function of $z$ of degree $Q,$ the simplest choice being to take $R=z^{Q}$.

This choice of $R=z^{Q}$ corresponds to a configuration which is axially symmetric. Note
 that for $Q=1$ the Skyrme field is spherically symmetric, but the Hopf projection breaks
 this symmetry so that only an axial symmetry remains. It is has been shown~\cite{KR} that spherical
 symmetry is not compatible with a non-zero Hopf charge, so the above axially symmetric 
configurations are the most symmetric candidates for soliton solutions.

The model (\ref{lag}) has a global $O(3)$ symmetry, but the choice of a vacuum
value ${\bf n}_\infty$ breaks this. There still remains a global $O(2)$ symmetry, which rotates
 the $n_1,n_2$ components. When we refer to a symmetry of a configuration, such as the axial 
symmetry mentioned above, it is not that the fields must be invariant under a corresponding 
spatial rotation, but rather that the effect of such a rotation can be undone by acting with 
the global symmetry of the theory. This implies that both the $n_3$ component (which determines
 the position of the soliton) and the energy density are strictly invariant under the symmetry
 operation.

Let us now analyze the axially symmetric configurations, given by $R=z^{Q}$, in a little 
more detail. First of all, the position of the soliton is the preimage of the vector
 ${\bf n}=(0,0,-1)$, which by equation (\ref{n3}) requires
that $\sin^2 f=1$ and $v_3=0.$ Given the properties of the profile function then the first requirement
 determines a unique radial value, $r_0$, such that $f(r_0)=\pi/2.$ The second requirement,
 when combined with equation (\ref{vr}), implies that $\vert R\vert =1$, and hence
 $\vert z\vert  =1.$ The preimage of $n_3=-1$ is therefore the circle of radius
 $r_0$ in the equatorial plane $\theta=\pi/2.$ To verify that the Hopf charge
 is indeed $Q$ we need to consider the preimage of another point on the target
 sphere, which we
choose to be the vector ${\bf n}=(0,-1,0).$ Using equations (\ref{n1}-\ref{n3})
the preimage is determined by the relations
\be
2(1-v_3^2)\sin^2f=1, \ \ 
v_3v_1\sin f=v_2\cos f, \ \
2\sin f(v_3v_2\sin f+v_1\cos f)=-1.
\label{pre2}
\ee
To examine the linking number of this preimage with the circle identified above
we need to consider where this preimage cuts the plane $\theta=\pi/2.$ In this plane $z=e^{i\phi}$
 giving $R=e^{iQ\phi}$ which when substituted into equation (\ref{vr}) 
gives ${\bf v}=(\cos Q\phi,\sin Q\phi,0).$ Substituting this expression 
into (\ref{pre2}) we deduce the constraints
\be
2\sin^2f=1, \ \ \sin 2f \cos Q\phi=-1, \ \ \cos f \sin Q\phi=0.
\label{cp2}
\ee
The first of the above equations determines exactly two radii, $r_\pm$, given
by $f(r_+)=\pi/4$ and $f(r_-)=3\pi/4,$ which from the properties of the profile function satisfy
 the inequalities $r_+>r_0>r_-.$ For the outer crossings, $r=r_+$,
the constraints (\ref{cp2}) reduce to 
$\cos Q\phi=-1$, which has exactly $Q$ solutions $\phi=(2p+1)\pi/Q$,
 $p=0,1,..,Q-1$, whereas
for the inner crossings at $r=r_-$, constraints (\ref{cp2}) reduce to 
$\cos Q\phi=1$, which also has $Q$ solutions $\phi=2p\pi/Q$, 
$p=0,1,..,Q-1$. By continuity these inner and outer crossings are smoothly 
connected together and thus this preimage is linked with the preimage of the vector 
${\bf n}=(0,0,-1)$ exactly $Q$ times. Note also that the inner and outer
 crossings are equally spaced in the angular direction around the position of the soliton. 
This completes the verification that the Hopf charge is $Q.$

Using the above axially symmetric initial conditions we have relaxed the 
configurations to find static solutions, which we shall discuss later. 
Of course, since we impose axial symmetry
from the start then all the resulting solutions preserve this axial symmetry.
However, as in the case of Skyrmions, we expect that most of these solutions
will be saddle points of the energy functional and be unstable to perturbations
which break the axial symmetry. To investigate this issue we require initial conditions
 which are perturbations of the ones we have constructed so far, so that the symmetry is
 broken. One possibility is to replace the rational map $R=z^{Q}$ by a more general
 degree $Q$ map and this often leads to similar results as those discussed below.
However it is also possible to create configurations with discrete symmetries,
which prevent the true minimum energy solution from being found, or if the rational map
is too generic it can lead to high energy configurations which break up into small
charge clusters. Given these problems we find that adding small
 perturbations to the axially symmetric configurations is more satisfactory.
Take the function $R$ to have the following form,
\be
R=z^{Q}\left[1+a\cos\left(\frac{m\phi^2}{2\pi}\right)\right]\,,
\label{toridef}
\ee
where $a\in[0,1)$ is a constant amplitude and $m$ is a constant integer.
Recall that $\phi$ is the phase of the angular variable $z$ so that $R$ is no
longer a meromorphic function of $z.$ By considering small deformations from this limit it is clear 
that the Hopf charge
 of this ansatz is still $Q$, though the parameter $a$ has to be restricted to $\vert a\vert<1$, 
so that the term multiplying $z^{Q}$ is non-zero. To see the physical significance of
 the deformation we can look at the position of the soliton. As before this is a closed loop 
lying a distance $r_0$ from the origin, but now the loop is not a circle in the plane
 $\theta=\pi/2.$ The angular distribution of the loop is given by the equation,
\be
\theta=2\mbox{tan}^{-1}\left[\left(1+a\cos\left(\frac{m\phi^2}{2\pi}\right)\right)^{-1/Q}
\right].
\ee
Thus as $\phi$ varies, that is, we move around the loop, it dips below and then rises
 above the equatorial plane $\theta=\pi/2.$ In other words, what was previously a circular
 loop now has wiggles with the amplitude being determined by the parameter $a$ and
 the number of wiggles controlled by the integer $m.$ Note that we use a quadratic rather than 
linear dependence on $\phi$ in the cosine argument  to ensure that the symmetry of the
 configuration is completely broken and that no cyclic subgroup remains. These are the
 initial conditions that we shall use to relax to the minima in the next section, with typical
 parameter values being
 $a=0.5$ and $m=50.$

\section{Relaxed soliton solutions}\news

In this section we describe the results of our relaxation computations using the initial
 conditions discussed in the previous section. We shall consider solitons of charges one
 to eight and display our results by plotting several interesting quantities. The first 
is the preimage of the vector $(0,0,-1)$ which defines the position of the soliton. In 
reality it is difficult to compute the locus of this preimage in the discretized domain 
and hence we will plot isosurfaces of the vector $(0,0,-1+\epsilon)$, where $\epsilon\approx 0.2$
 is small. This allows us to easily visualize the solitons position as a line-like solid rather
 than a single line. We will also explicitly display the linking number, thereby verifying
 the Hopf charge, by plotting, in a similar way to the position, the loci of two independent 
points $(0,-1,0)$ and $(0,0,-1)$. Finally, we also plot the isosurfaces of energy density,
 sometimes superposed with the position of the soliton so as to characterize the maxima of
 the energy density. The position of the initial conditions for each charge is shown in fig.~\initial.ps

It is possible to compute the total energy of the discretized configurations by integrating the 
local energy density over the box. This will in general under estimate the energy of the 
configuration since the box is finite. In the case of Skyrmions it was possible to make a 
better estimate of the total energy by dividing by the topological charge of the discretized 
configuration. It was claimed that this allows one to estimate the energies to within $1\%$.
 However, this is not possible here since we have no expression for the local charge density.
 Hence, there will be systematic uncertainties in the computations which could be upto $5\%$,
but comparisons between the energies computed for different configurations should be more accurate. 

\subsection{Low charge : $Q=1,2,3$}\news

First, let us discuss in detail the solutions for $Q=1,2,3$ which were studied in
 refs.~\cite{FN,GH} In each case we investigated an exactly axially symmetric solution
 and also the inclusion of non-symmetric perturbations given by (\ref{toridef}).

For $Q=1$ the solutions relaxed from the symmetric and non-symmetric initial
 conditions are identical, with the symmetric case remaining almost unchanged except 
for a slight rescaling; the results from the non-symmetric initial conditions being
 presented in fig.~\one.ps, showing the position of the soliton, the linking 
structure, the energy density profile, and a comparison between the energy density and
 the position. The line representing the position of the soliton is a simple circle 
and it is clear that the preimages of $(0,-1,0)$ and $(0,0,-1)$ are linked exactly 
once, hence the relaxed solution we have computed has maintained the initial Hopf charge
 $Q=1.$ The surface of constant energy density has an ellipsoid shape and is not
 spherically symmetric. When compared to the position of the soliton the energy density
 is seen to be localized well inside the position, with the maximum value being at the 
origin, as found in~\cite{GH}, and not on a torus surrounding the string as found in~\cite{FN}.
 The energies of the two configurations, which are presented in table~1, are also equal
 (modulo our earlier discussion of the difficulties in computing the energy), the precise
 value being more than double the energy bound for $Q$, in good agreement with the 
work of Gladikowski and Hellmund~\cite{GH}. The fact that symmetric initial conditions
 remain almost unchanged suggests that with an appropriate choice of the profile function,
 the soliton we obtain is reasonably well approximated by the axially symmetric initial
 conditions with $R=z.$

Upon relaxation the behaviour of the $Q=2$ initial conditions, both symmetric and
 non-symmetric, is very similar to that of $Q=1$. Once again we present the results 
from the non-symmetric initial conditions, this time in fig.~\two.ps. As
 for $Q=1$, the position of the soliton is a simple torus, but now the 
energy density is also toroidal, although the line of maximum energy density is
 inside that of the position. From table 1, it is easy to see that the total energy 
computed for $Q=2$ is much less than twice that for $Q=1$, suggesting
 that the 2-soliton configuration is tightly bound relative to decomposition into
 two 1-solitons.  Once again, the linking structure confirms the Hopf charge, the 
preimages being doubly linked, and the under and over crossings of the two preimages 
are equally distributed in the angular direction, as described in the previous section
 for the rational map ansatz. Our results agree well with those of both ref.\cite{FN}
 and ref.\cite{GH} for this charge. Once again, similarity of the configurations relaxed
 from symmetric and non-symmetric initial conditions, suggest that the minimum energy
 configuration is well represented by a map with $R=z^2.$ 

The results of relaxing non-symmetric initial conditions for $Q=3$ are displayed
 in fig.~\three.ps, while the symmetric initial conditions once again relax to a
 rescaled axially symmetric configuration. Since 
the two are not the same it appears that the minimum energy configuration is not a torus.
 The position of the soliton is a closed loop, but this time it is twisted. Thus, it 
appears that the twisting of the field lines required to generate this Hopf charge makes
 it energetically favourable for the position of the soliton to twist also. It should be
 noted that the twist in the 3-soliton found after relaxation is not related to the wiggles
 generated  in the initial conditions, which are of much smaller amplitude and greater in
 number. The energy density profile of this solution has a rather unusual shape, appearing
 to be something very similar to a pretzel, but with two holes and a twist. The total energy
 of the configuration created from non-symmetric initial conditions is slightly lower than
 the toroidal configuration, but given the uncertainties in the computing the energies we 
have eluded to earlier, this cannot be totally convincing. However, we have tried a number 
of different random initial configurations, all of which relax quickly to the same `twisted torus'
 and therefore we believe that we have created a non-axially symmetric minimum energy configuration.
 Gladikowski and Hellmund~\cite{GH} did not attempt to construct a $Q=3$ configuration,
 but if they had using their assumption of axial symmetry they would have found the torus, rather
 than the twisted torus. Faddeev and Niemi \cite{FN} presented some preliminary numerical
 evidence that there is a stable trefoil knot configuration at this charge. Given the results
 presented here we believe this to be unlikely, although we will discuss in section 6, 
our attempts to construct such a configuration.

\subsection{Higher charge : $Q=4,5,6,7,8$}

When we relax non-symmetric initial conditions for $Q=4$ and $Q=5$, 
the solutions are very different to those created from symmetric ones. As for $Q=3$
 the minimum energy configurations appear to be located on twisted closed loops and the energy
 density distributions are twisted pretzels, which have three and four holes respectively;
the computed energies of these discretized minimum energy
 configurations being smaller than the 
corresponding tori (see table 1). The positions, linking structures and energy density profiles of these
 solitons are summarized with all the others in figs.~\pos.ps , \hopf.ps and 
\ene.ps respectively. 

Above $Q=5$ there appears to be a dramatic change in the structure of the soliton solutions,
as can be seen from figs.~\pos.ps , \hopf.ps and 
\ene.ps . At $Q=6$ we see that the position of the soliton
is no longer a single connected loop, but consists of two disjoint loops, which moreover
are linked. Thus, remarkably we have found that a linked loop has emerged from a rather general asymmetric
initial condition consisting of a single loop. Note that the linking
number of the position is one, which should not be confused with the linking number
$Q$ which determines the soliton charge and corresponds to the linking number of
two loops obtained as the preimages of two distinct points on the target sphere.
However, the fact that the position of the soliton, which we recall is itself a preimage,
is disconnected and linked means that the counting of the Hopf charge $Q$ is now
a subtle matter. In fact a careful examination of the linking of field lines
in fig.~\hopf.ps reveals that the $Q=6$ configuration resembles two
linked $Q=2$ solitons. However, the Hopf charge is not simply additive in the
case of a link since when a field line passes through the intersection 
of the link it should be counted twice. This happens exactly once for each of 
the two $Q=2$ solitons and hence rather than being a total charge of $Q=4$, it is indeed
 a $Q=6$ soliton. Note that the energy density does not have the form of two linked loops,
 but rather is concentrated mainly in the region where the
two position loops are linked. It seems that the interaction between the
loops contains some of the energy of the soliton, and it can reach this lower
energy state by smearing out the energy over a small region, rather than
localizing it around the positions of the solitons. This is, of course,
consistent with our earlier findings at lower charge where the $Q=1$ soliton
has its energy smeared out inside the position loop and even the $Q=2$ soliton
energy is maximal on a loop inside the position loop.
We shall see that the key to understanding the formation of this linked soliton 
is the property of string reconnection discussed in section 5. Therefore,
the detailed formation of the structure of the $Q=6$ link, and the other exotic
solitons introduced below, will be left until section 6.

From fig.~\pos.ps it is clear that the position of the $Q=7$ soliton has the
structure of a trefoil knot. It was suggested by Faddeev and Niemi \cite{FN}
that knotted solitons might exist in this model, although it was conjectured that the
 $Q=3$ soliton had the structure of a trefoil knot. They attempted
to justify this numerically by imposing a trefoil knot structure in their
initial conditions, but we have already shown using our much more general numerical algorithm
 that this is not the case. Here, we have a much more satisfactory situation, in that a trefoil
 knot --- the simplest knot configuration --- has emerged naturally from general asymmetric
 initial conditions. This not only shows that knotted solitons indeed occur in this model,
 but suggests that they are important configurations which arise naturally and do not require 
any fine-tuning of the initial conditions. We should note that only the locus of the position 
of the soliton is a trefoil, and that the preimage of other points (for example, (0,-1,0) as 
shown in fig.~\hopf.ps) will  not be. Neither is the energy density profile, which was part
 of the argument in favour of knotted solitons in ref.~\cite{FN}. In fact, the energy is
 smeared out inside the knot.

At $Q=8$ we find yet another new phenomenon: the position of the soliton is again two linked loops,
 but in contrast to the link at $Q=6$, these links
themselves have a higher linking number. In fact the position loops are doubly
linked, which makes the task of counting the Hopf charge even more difficult. It appears that the 
solution comprizes of two $Q=2$ solitons doubly linked, the missing Hopf charge coming again from 
the double linking of the position.

\subsection{Soliton properties and energy minimization} 

In table 1 we give the energy values of the solitons computed from
the general asymmetric initial conditions discussed earlier. For comparison we
also list the energies of the axially symmetric tori solutions (whose relative sizes
 are shown in fig.~\tori.ps for charges one to six) obtained by imposing
axial symmetry in the initial conditions by using a rational map of the form $R=z^Q$.
Recall that for $Q=1$ and $Q=2$ we found that the relaxed solitons regained their
axial symmetry even when it was initially broken by a perturbation, so the energy
values for these solitons should be equal to those of the tori. The minute differences
seen in the first two rows of table 1 are the result of our numerical computation
of the total energy, which as discussed earlier is susceptible to errors.

\medskip
\begin{center}
\begin{tabular}{|c|c|c|}
 \hline 
$Q$ & Soliton Energy & Torus Energy\\
\hline 
1 & 504  &  505  \\
2 & 835  &  836  \\
3 & 1157 &  1181 \\
4 & 1486 &  1542 \\
5 & 1808 &  1974 \\
6 & 1981 &  2361 \\
7 & 2210 &  2600 \\
8 & 2447 &  3050 \\
\hline
\end{tabular}
\end{center}
{\bf Table 1 } : Energy of the relaxed soliton and torus solutions for charges one to eight.
 As discussed in the text, the absolute values are likely to be systematic under estimates,
 but comparisons between configurations are likely to be qualitatively correct.
\medskip

There are two obvious conclusions to be drawn from table 1. The first is that the multi-solitons
 are all easily bound against the break-up into smaller soliton clusters, which one would expect
 if the energy of the solutions is close to having the $Q^{3/4}$ dependence suggested by the energy 
bound. The second is that the axially symmetric torus solutions rapidly increase in energy when 
compared to the soliton solutions we have found, clearly
demonstrating the lack of axial symmetry in the general charge $Q$ solution.

Now that we have computed the energies of the solitons upto charge eight, it is possible to begin to
 investigate the energy bound (\ref{bound}) and
the growth of the soliton energy with increasing charge.
As mentioned earlier the energy of the $Q=1$ soliton is much larger than the
bound (\ref{bound}) with the given value of the constant $c$, and hence it is
of little practical use. However, since the bound is derived under some very general assumptions, 
the coefficient computed may be artificially small. If the fractional power of  the energy bound is correct
 then it may be possible to empirically, and within the systematic errors in computing the energy
 already discussed, improve upon the coefficient. Let us assume that
an energy bound of the form (\ref{bound}) exists, but with a new value for the
constant $c.$ Clearly the tightest bound which could be obtained would be with
the value $c=E_1$, where $E_1$ is the energy of the $Q=1$ soliton. We therefore
choose to compare our energy values with this \lq optimal bound\rq, that is,
 $E_1Q^{3/4}.$ In fig.~\energy_graph.ps we plot the soliton energy (crosses)
against this \lq optimal bound\rq\ (dashed line). It can be seen that the soliton
energies lie very close to this curve, suggesting that a true bound exists which
is very close to this one and moreover that the bound would be very tight.
The plot also supports the fractional power growth of the energy, which is an unusual
feature, and clearly our results are not consistent with the typical linear growth, $E_1Q$ 
(dotted line) common in many topological soliton models. Also shown on the plot
are the energies of the tori saddle points (diamonds), which are also well below
the linear growth curve, thereby indicating that they too are bound against
a break-up into $Q$ well separated 1-solitons.

From our results we can speculate on some qualitative aspects of an energy minimization
 principle which leads to the interesting structures we have found.  A similar approach was 
taken for Skyrmions~\cite{BS2,BS3} and it was deduced that a simple mechanical principle was 
at work. Since the solitons here are line-like, the total energy is $E=\mu L$, where $\mu$ is 
the energy per unit length --- naively assumed to be universal --- and $L$ is the length of string. 
Therefore, a simple mechanical analogy in this case would be that the length of string must be minimized
 subject to some constraint, which must be related to the imposition of the correct topological charge,
 with the relative linking of two preimages requiring a certain amount of gradient energy. For $Q=1$
 and $Q=2$, one could easily imagine that such a principle requires that the solutions be toroidal,
 with the twists distributed uniformly. It is clear from the linking structures seen for $Q=3,4,5$ 
that it is possible for the links to be packed closer together with discrete symmetries, hence reducing
 the length of string required for the position. What happens for the higher charges is less well defined,
 but it is clear that having extra links in the position itself can reduce the number of links in each 
of the individual solitons, as well as localizing the energy. This would suggest that the position 
itself prefers to be linked as much as possible in order for the constituent parts to 
have the smallest possible 
charge. This idea is borne out for $Q=8$, where two doubly linked $Q=2$ solitons are formed in preference
 to two singly linked $Q=3$ solitons. 

The clear oddity is $Q=7$ which is the first structure where the position is actually knotted.
 The reason why being knotted is preferred over being  linked in this case is probably a question 
of symmetry. If one were to split $Q=7$ into two singly linked structures, then removing two for the 
link would require a total of five to be shared out between them. Clearly, this cannot be done 
symmetrically and hence a knotted structure is energetically preferred. We have observed some
 evidence for this during the numerical relaxation of different initial conditions. Often a solution
 would become trapped in a linked structure (either $Q=1$ linked with $Q=4$, or $Q=2$ linked with $Q=3$)
 with energy substantially higher than the computed minimum.
 
Obviously more examples of higher charge solitons will be required before a detailed understanding of
 the complete energy minimization principle can be found, but we expect that these features will play
 a prominent role. One thing that is transparent in this discussion is that as the charge increases,
 the number of possibilities for linking proliferates, making the prediction of an energetically preferred
 configuration increasingly difficult. If sufficient interest is generated, many more hours of
 supercomputer time could be spent in search of a more general principle!

\section{Reconnection of Skyrme Strings}\news

As mentioned in section 2, the initial motivation for the model under investigation here was 
the observation by Faddeev~\cite{Fa} that a planar soliton with an internal degree of freedom
 could be embedded in three-dimensional space in such a way that the internal twisting of the 
soliton produces a topological obstruction to its decay. This lead Faddeev and Niemi to the 
conjecture that all knotted configurations will exist in this model and indeed we have shown 
that this may be possible, although not quite in the way that they suggested. In order to reflect
 on this very wide ranging possibility it is important to understand the processes by which parts of the 
string may interact.

There are a number of systems in (2+1)-dimensions, such as
the Abelian Higgs model, which possess topological solitons
whose energy density is localized around a point in space,
given by the position of the soliton. If such models are
considered in (3+1)-dimensions then the soliton can be 
trivially embedded into the extra dimension to produce
an infinite string-like object, whose energy density is
localized along a line. Cosmic strings~\cite{VS} are the 
best studied examples of this procedure, where the planar
soliton is a vortex. By embedding the vortex
along a closed curve, rather than a straight line, the ends
of the string can be joined to produce a closed string of finite
length. However, such closed strings tend to be unstable
configurations and will collapse when considered as dynamical
structures (see, for example, ref.\cite{BS}). The reason a vortex in the plane is stable is
due to the existence of a topological conservation law, but when
the vortex is embedded to produce a closed string, this
topological stability is lost. In the case of a circular loop this can be 
understood by realizing that in a plane perpendicular to the loop, one has 
effectively a vortex/anti-vortex pair.

The planar solitons of relevance to the more complicated model discussed in this paper are
 known as Baby Skyrmions~\cite{PSZ} and arise in the model in (2+1)-dimensions described
 by the Lagrangian density,
\be
{\cal L}=\partial_\mu \bn\cdot \partial^\mu \bn
-\frac{1}{2}(\partial_\mu \bn \times \partial_\nu \bn)\cdot
(\partial^\mu \bn \times \partial^\nu \bn)-\beta^2(1-n_3)
\,.
\label{lagbs}
\ee
Note that this Lagrangian is the planar version of (\ref{lag})
except that an additional potential term, with constant coefficient $\beta^2$,
has been added. This additional term stabilizes the solitons to the radial scaling 
of Derricks theorem, since in two space dimensions
the pure sigma model (the first term of (\ref{lagbs})) is scale invariant. In fact any
 term containing no derivatives, such as a potential,  would suffice for this purpose
 and the particular choice here was chosen by the authors in ref.\cite{PSZ} by analogy to
  the pion mass term of the $SU(2)$ Skyrme model. It has the added advantage that the
 Baby Skyrmions are then exponentially localized. Although the details of the Baby Skyrmions
 are sensitive to the choice of a potential the features we are concerned with in this section
 are topological and hold for any choice of potential term, including zero.

Although a potential term is not required in (3+1)-dimensions one could be
included, such as the one given above for Baby Skyrmions, though we have chosen
not to include one at this stage in order to make the model as simple as possible.
Some investigations on how this can affect the  properties of the solitons of charge 
one and two have been undertaken~\cite{GH}.

Since the field of the model
takes values in a two-sphere then in the planar case the
relevant quantity is that $\pi_2(S^2)=\Z$,
so that a Baby Skyrmion is a topological soliton having
a non-zero winding number.
A single static Baby Skyrmion  has the hedgehog form
\be
{\bf n}=(\sin g\cos\theta,\sin g\sin\theta,\cos g)
\label{hh}
\ee
where $(\rho,\theta)$ are polar coordinates in the plane and $g(\rho)$
is a profile function satisfying the boundary conditions $g(0)=\pi$ and
$g(\infty)=0.$ The internal phase of this soliton is the freedom to add
a constant angle to $\theta$, which rotates the components $(n_1,n_2).$ 
As in the three-dimensional case, the position of the Baby Skyrmion is the 
preimage of the point $(0,0,-1)$, which in the above example has been chosen to be the origin.

If this Baby Skyrmion is now placed into the (3+1)-dimensional model
(\ref{lag}) by embedding it along a closed curve, then the curve traced
out by the position of the Baby Skyrmion will be a closed loop and will
be the preimage of the point $(0,0,-1).$ If, in addition, the internal phase
of the Baby Skyrmion changes by a total angle of $2\pi Q$ as it moves around the loop
then this generates a finite energy configuration with Hopf charge $Q.$

In order to comment on the conjecture that knotted solitons are inevitable in this model,
 we wish to study the  process of loop intersection
and this is most easily examined without the additional complications of curvature
effects. We therefore choose to analyze this feature by considering infinite straight
strings, rather than closed loops. We define a Skyrme string as a Baby Skyrmion
solution (\ref{hh}) which is trivially embedded into the (3+1)-dimensional version
of (\ref{lagbs}) by making the fields independent of the third Cartesian direction. 
Clearly the choice of the $x_3$-direction is immaterial here and a Skyrme string exists as a
solution for any choice of embedding direction.

These Skyrme strings are the analogues of infinite cosmic strings, but where the vortex 
of the Abelian Higgs model is replaced by the Baby Skyrmion. It is well understood that
the intersection of two cosmic strings always leads to the phenomenon of string
 reconnection~\cite{VS,MSB}, where the two strings break at the region of intersection and then reconnect
after a change of partners. This process is purely topological and can be seen as a
mixture of the right angle scattering of a vortex-vortex interaction with the annihilation
of a vortex-anti-vortex pair~\cite{MSB}. Baby Skyrmions also scatter at right angles and annihilate
if they have opposite topological charge~\cite{PSuZ} so we may expect that Skyrme strings
also exhibit reconnection.

Using numerical simulations have investigated Skyrme string reconnection for a wide range of parameters.
We evolve the full equations of motion which follow from the (3+1)-dimensional version of
(\ref{lagbs}), with the parameter value $\beta=0.45.$ We take a product ansatz to form
the initial conditions consisting of two infinite straight Skyrme strings lying in 
the planes $x_2=\pm a$, forming
 angles $\pm\Theta/2$ with the $x_3$ axis and each Lorentz boosted with speed $v$ towards
each other. The results are similar for varying values of $a$ and $v$, but typical values 
used are $a=1$ and $v=0.2$. Fig.~\recon.ps displays energy density isosurfaces as time evolves for
three particular initial conditions corresponding to initial relative string angles of 
$\Theta=\pi/6,\pi/2,5\pi/6.$ In each case reconnection takes place, as expected.
Note that the initial conditions in fig.~\recon.ps.a and fig.~\recon.ps.c look very similar, but that in
the first case the strings are almost parallel, whereas in the last case they are almost
anti-parallel. This difference becomes apparent in the time evolution when the choice of
reconnection partners is made, with the anti-parallel simulation being a much more violent
process, producing more radiation due to the large soliton-anti-soliton
component of the interaction. Note that if both strings were exactly parallel then this would
be nothing more than  Baby Skyrmion right angle scattering, and if they were exactly anti-parallel 
then both strings would annihilate, as it would correspond to a planar soliton-anti-soliton interaction.

We have performed several simulations with varying values of $a,v,\Theta$ and $\beta$
and always found that reconnection takes place in the manner described above.
Even for $\beta=0$, in which case the model is exactly (\ref{lag}), 
 the results are very similar, although the scale of the strings may now vary,
as there is now no potential term to provide a stable size for the Baby Skyrmions.
It is likely that other potential terms which respect the topology will influence the 
interaction between strings,
such as whether they attract or repel, but the process of reconnection is topological
and therefore will occur for all sensible choices of the potential. We have also varied the relative
internal phases of the Skyrme strings and verified that reconnection still continues.
Note that since the force between two Baby Skyrmions depends upon their 
internal phases then there are interesting possibilities for the dynamics of our Skyrme
strings if the internal phases are varied along the length of the string. In particular
it might be possible that the strings could be made to wrap around each other.
A more detailed investigation of reconnection in this model is in progress.

We have demonstrated, therefore, that reconnection takes place for infinite parallel strings 
 and we are now in a position to discuss the possibility of the formation of linked and knotted 
soliton structures. The fact that reconnection can take  place illustrates that the stability of
 links and knots is by no means inevitable in this model and hence the exotic linked and knotted
 configurations already discussed appear to be a consequence of subtle cancellations between the
 interactions of different
 pieces of string.

\section{Links and Knots}\news

In fig.~\sixmov.ps we display the time evolution of the position of the soliton 
for the $Q=6$ configuration during the relaxation process, 
from the initial conditions fig.~\sixmov.ps.1 to the final linked loops fig.~\sixmov.ps.8. 
The initial condition is a small deformation of a  circular loop which  very quickly develops a
 number of large twists. In fact, these twists are so severe
that by fig.~\sixmov.ps.5 the loop has come close to self-intersection. Intersection
does then take place with the reconnection process proceeding  as described in the previous
section, leading to the formation of two linked loops. Thus, we see that the combination
of the twisting with the reconnection property of the strings has led
rather naturally to a link.

In fig.~\sevenmov.ps we display the time evolution of the position of the soliton for the 
$Q=7$ configuration and we see that initially it proceeds in a similar manner to the $Q=6$ soliton.
 The twisting and reconnection again result in two linked loops fig.~\sixmov.ps.4, but this time
 since the Hopf charge is odd it is impossible for two linked loops to equally distribute the Hopf 
charge and hence the energy. This results in an instability of this link which it corrects by performing 
yet another reconnection, thereby leading to the final state having the structure of a trefoil knot.
 Since the link and the trefoil only differ by a single reconnection it is apparent that whether 
links or knots result in the final state may be a delicate matter. It is also likely that at high
 charge there may be several meta-stable states corresponding to various linked and knotted configurations.

The evolution of the $Q=8$ soliton, shown in fig.~\eightmov.ps, follows a similar 
pattern of twisting and in this case a number of reconnection events. This leads finally
 to a configuration which comprizes of two $Q=2$ solitons that  are doubly linked. This
 configuration appears to form in preference to singly linked and knotted configurations.
 The instability of the toroidal configuration to twists seen for $Q=6,7,8$ is further
 evidence that the axially symmetric configurations are unstable.

Finally, we should comment on our attempts to construct a trefoil knot with $Q=3$ as
 was suggested in ref.~\cite{FN}. Clearly, such a configuration is possible, since it will
 have the correct linking number if the preimage of each point on the target sphere is a
 trefoil. However, will it be stable? In order to construct initial conditions for such a
 configuration, one must modify the rational map based approach discussed in section 2. The
 precise details of how to do this are presented in ref.~\cite{BSproc}, but once the initial
 solution is relaxed it appears to settle down to a configuration which comprizes of two 
linked loops.
Note that even the numerical requirement of placing the trefoil in a finite box introduces
perturbations which after some time allow the symmetry of the trefoil knot to be broken.
 The initial energy of the trefoil is around three times that of the relaxed 
minimum energy configuration, while that of the final relaxed configuration is twice the minimum.
 The fact that the solution does not relax down to the minimum suggests that there is some
remaining symmetry from the initial conditions we use and hence the two linked loops are a symmetric 
saddle point. We should note that the ansatz we use for the initial trefoil is by no means 
unique and hence it might be possible for us to construct a stable trefoil configuration using
 more symmetry. However, its energy is likely to be far too high for it to be relevant in any
 physical application of this theory. It appears that the energy minimization principle
 discussed in section 4 
is in action here. Obviously, a much larger piece of string (and hence energy) is required
 for the links to be arranged in the trefoil shape at this charge.

\section{Conclusions}\news

Using numerical simulations we have investigated solitons, stabilized by the Hopf invariant,
 in a modified $O(3)$ sigma model. Our results follow on from those of refs.~\cite{FN,GH} 
for $Q=1,2$, confirming the results of ref.~\cite{GH} where differences occur. Using our
 more general numerical algorithm and effectively random initial conditions, we have been 
able to investigate higher charges, illustrating that for $Q=3,4,5$ the solitons have the 
form of twisted tori, rather than knots as suggested in ref.~\cite{FN}. However, the basic
 premise that knots and other exotic linked solitons can occur in this model is confirmed
 for $Q=6,7,8$ on the basis of these numerical computations. We have attempted to construct
 a qualitative energy minimization principle
 based on the analogy to the mechanics of a piece of string. We propose that the length of
 string is minimized subject to the requirement that there are sufficient twists for a 
particular Hopf charge. It appears that the position of the soliton will link with itself
 once there is sufficient string for sensible daughter links to be created and that if this
 cannot be done symmetrically, then a knot will be formed. It is obvious that as the charge 
increases, the number of possibilities for the linking structure will increase rapidly and
 that reconnection will play an important role in this. Hence, many more hours of CPU time
 will be required to pin down a more quantitative principle.

In a similar study of Skyrmions \cite{BS2} we were able to demonstrate that 
interesting structures appeared in the soliton solutions and suggest an 
energy minimization principle responsible for their formation. Subsequently
these numerical results were supported by an approximate analytical approach,
involving an ansatz based on rational maps between Riemann spheres \cite{HMS}.
It would be desirable to construct a similar approximate technique for the 
solitons computed in this paper, and perhaps the rational map generated initial
conditions or the more general rational map ansatz introduced in ref.\cite{BSproc}
may be a useful starting point.

In the introduction we mentioned that there are a number of possible physical
 manifestations of this model. Now that we have made this preliminary investigation of the
 Hopf stabilized solitons in this model, we can now turn our attention to these.
Aspects such as the dynamics and scattering of these solitons, both in relativistic and dissipative
versions of the model, will then need to be investigated in great detail and preliminary
investigations are already underway.\\

\noindent {\sl Notes added.}

\smallskip
1. A suggested improved value
for the constant $c$ in the energy bound has recently been proposed
 -- R.S. Ward, {\sl Durham preprint DTP-98/55}.
This value is approximately double the value in the known bound and hence is in good agreement
with the results in this paper.

2. Some of our findings on the structure of the soliton solutions (for example the twisted
torus at charge three) have recently been confirmed by numerical relaxation calculations
which use very different numerical algorithms and initial conditions from the ones employed here --
J. Hietarinta and P. Salo, hep-th/9811053.\\

\section*{Acknowledgments}

\noindent Many thanks to Ludwig Faddeev, Jens Gladikowski, Meik Hellmund, Conor Houghton, Patrick Irwin,
 Nick Manton, Antti Niemi, Miguel Ortiz, Richard Ward, and Wojtek Zakrzewski
 for useful discussions. We would like to acknowledge the use of the SGI Origin 2000
 and Power Challenge at DAMTP in Cambridge supported by the HEFCE, SGI, PPARC, the 
Cambridge Relativity rolling grant and EPSRC Applied Mathematics Initiative grant GR/K50641. 
The work of PMS is supported by an EPSRC grant GR/L88320 and Advanced Fellowship AF/98/0443 
and that of RAB is supported by Trinity College.
 We would like to thank Paul Shellard for his tireless efforts to provide sufficient
 computational resources for projects such as this.

\def\jnl#1#2#3#4#5#6{{#1 [#2], {\it #4\/} {\bf #5}, #6.} }
\def\jnltwo#1#2#3#4#5#6#7#8{{#1 [#2], {\it #4\/} {\bf #5}, #6;{\bf #7} #8.} }
\def\prep#1#2#3#4{{#1 [#2], #4.} } 
\def\proc#1#2#3#4#5#6{{#1 [#2], in {\it #4\/}, #5, eds.\ (#6).} }
\def\book#1#2#3#4{{#1 [#2], {\it #3\/} (#4).} }
\def\jnlerr#1#2#3#4#5#6#7#8{{#1 [#2], {\it #4\/} {\bf #5}, #6.
{Erratum:} {\it #4\/} {\bf #7}, #8.} }
\def\prl{Phys.\ Rev.\ Lett.}
\def\pr{Phys.\ Rev.}
\def\pl{Phys.\ Lett.}
\def\np{Nucl.\ Phys.}
\def\prp{Phys.\ Rep.}
\def\rmp{Rev.\ Mod.\ Phys.}
\def\cmp{Comm.\ Math.\ Phys.}
\def\mpl{Mod.\ Phys.\ Lett.}
\def\apj{Ap.\ J.}
\def\apjl{Ap.\ J.\ Lett.}
\def\aap{Astron.\ Ap.}
\def\cqg{Class.\ Quant.\ Grav.} 
\def\grg{Gen.\ Rel.\ Grav.}
\def\mn{M.$\,$N.$\,$R.$\,$A.$\,$S.}
\def\ptp{Prog.\ Theor.\ Phys.}
\def\jetp{Sov.\ Phys.\ JETP}
\def\jetpl{JETP Lett.}
\def\jmp{J.\ Math.\ Phys.}
\def\zpc{Z.\ Phys.\ C}
\def\cupress{Cambridge University Press}
\def\oup{Oxford University Press}
\def\pup{Princeton University Press}
\def\wss{World Scientific, Singapore}

\newpage

\section*{Figure Captions} 

\noindent 

\noindent Fig.~\sketch_hopf.eps.
A sketch showing two loops corresponding to the  preimages of two points on the target sphere.
 The loops are linked exactly once, indicating that the configuration has Hopf charge $Q=1$.\\

\noindent Fig.~\initial.ps
An isosurface plot of the locus of position for the initial 
conditions used in the relaxation process for $Q=1$ to 8. Each one is close to being a circle,
 but it has randomly placed wiggles superimposed on it.\\

\noindent Fig.~\one.ps.
Isosurface plots displaying the locus of the position, the energy density, linking structure between 
two independent points on the target sphere and
a comparison between the position and energy density, for the $Q=1$ soliton.
 Notice that the linking number is indeed one and that the energy density is not a torus, but rather
its maximum is a point inside the locus of the position.\\

\noindent Fig.~\two.ps.
The same quantities as in fig.~\one.ps, but for the $Q=2$ soliton. Notice that both locus of the 
position and the energy density are both toroidal, but that the energy density is peaked inside 
the position.\\

\noindent Fig.~\three.ps.
As fig.~\one.ps but for the $Q=3$ soliton. Clearly, the position of the solition is now not axially 
symmetric.\\

\noindent Fig.~\pos.ps.
Isosurface plots showing the position of the soliton for $Q=1$ to $Q=8$.\\

\noindent Fig.~\hopf.ps.
Isosurface plots showing the linking number of the soliton for $Q=1$ to $Q=8$.\\

\noindent Fig.~\ene.ps.
Isosurface plots showing the energy density of the soliton for $Q=1$ to $Q=8$.\\
 
\noindent Fig.~\tori.ps 
Isosurface plots showing the locus of the position after the relaxation of the symmetric (toroidal)
 initial conditions.\\

\noindent Fig.~\energy_graph.ps.
A plot of the soliton energy (crosses) and energy of the torus solutions (diamonds) for Hopf charge 
from $Q=1$ to $Q=8.$ Also shown is a linear growth of energy $E_1Q$ (dotted line), and a fractional
 power growth in energy $E_1Q^{3/4}$(dashed line). Here $E_1$ is the energy of the $Q=1$ soliton. 
The plot clearly displays the fractional power growth of the soliton energy with Hopf charge $Q.$\\

\noindent Fig.~\recon.ps.
Energy density isosurfaces at increasing times for two reconnecting Skyrme strings. The three 
different sequences correspond to a relative angle between the strings of (a) $30^\circ$; (b)
 $90^\circ$; (c) $150^\circ.$\\

\noindent Fig.~\sixmov.ps.
Isosurfaces displaying the position of the soliton during the relaxation of a
$Q=6$ soliton.\\

\noindent Fig.~\sevenmov.ps.
Isosurfaces displaying the position of the soliton during the relaxation of a
$Q=7$ soliton.\\
 
\noindent Fig.~\eightmov.ps.
Isosurfaces displaying the position of the soliton during the relaxation of a
$Q=8$ soliton.\\


\begin{thebibliography}{99}

\bibitem{BS} 
\jnl{R.A. Battye and  E.P.S. Shellard}{1994}{}{\np}{B423}{260}

\bibitem{BS1} 
\jnl{R.A. Battye and P.M. Sutcliffe}{1997}{}{\pl}{391}{150}

\bibitem{BS2}
\jnl{R.A. Battye and P.M. Sutcliffe}{1997}{}{\prl}{79}{363}

\bibitem{BS3}
\prep{R.A. Battye and P.M. Sutcliffe}{1998}{}{DAMTP-1998-108}

\bibitem{BS5}
\prep{R.A. Battye and P.M. Sutcliffe}{1998}{}{hep-th/9808129}

\bibitem{BSproc} 
\proc{R.A. Battye and P.M. Sutcliffe}{1998}{}{Solitons: the CRM Monographs in
 Mathematical Physics}{R. B. MacKenzie, M. B. Paranjape and
W. J. Zakrzewski}{Springer-Verlag}
 
\bibitem{Fa} 
\prep{L. Faddeev}{1975}{}{Princeton preprint IAS-75-QS70}

\bibitem{FN} 
\jnl{L. Faddeev and A.J. Niemi}{1997}{}{Nature}{387}{58}\prep{L. Faddeev and A.J. Niemi}{1997}{}{hep-th/9705176}

\bibitem{FN2}
\prep{L. Faddeev and A.J. Niemi}{1997}{}{hep-th/9807069}

\bibitem{GH}
\jnl{J. Gladikowski and M. Hellmund}{1997}{}{\pr}{D56}{5194}

\bibitem{derrick}
\jnl{R. Hobart}{1963}{}{Proc. Roy. Soc. Lond.}{82}{201}\jnl{G. Derrick}{1964}{}{J. Math. Phys.}{5}{1252}

\bibitem{HMS}
\jnl{C.J. Houghton, N.S. Manton and P.M. Sutcliffe}{1998}{}{\np}{B510}{507}

\bibitem{KR}
\jnl{A. Kundu and Y.P. Rubakov}{1982}{}{J. Phys.}{A15}{269}\jnl{A.F. Vakulenko and L.V. Kapitanski}{1979}{}{Dokl. Akad. Nauk USSR}{246}{840}

\bibitem{Me}
\jnl{U.G. Meissner}{1985}{}{\pl}{154B}{190}

\bibitem{Pa} 
\proc{N. Papanicolaou}{1993}{}{Singularities in Fluids, 
Plasmas and Optics}{R.E. Caflisch and G.C. Papanicolao}{Kluwer Amsterdam}

\bibitem{PSZ} 
\jnl{B.M.A.G. Piette, B.J. Schroers and W.J. Zakrzewski}{1995}{}{\np}{B439}{205}

\bibitem{PSuZ} 
\jnl{B.M.A.G. Piette, P.M. Sutcliffe and  W.J. Zakrzewski}{1992}{}{Int. J. Mod. Phys.}{C3}{637}

\bibitem{MSB}
\prep{J.N. Moore, E.P.S. Shellard and R.A. Battye}{1998}{}{In preparation}

\bibitem{Sk}
\jnl{T.R.H. Skyrme}{1962}{}{\np}{31}{556}

\bibitem{monopole}
\jnl{P.M. Sutcliffe}{1997}{}{Int. J. Mod. Phys.}{A12}{4663}

\bibitem{VS} 
\book{A. Vilenkin and E.P.S. Shellard}{1994}{Cosmic
strings and other topological defects}{\cupress}

\end{thebibliography}
\end{document}